 \documentclass[conference]{IEEEtran}

\IEEEoverridecommandlockouts
\usepackage[T1]{fontenc}
\usepackage[latin9]{inputenc}
\usepackage{color}
\usepackage{graphicx}
\usepackage{epsfig}
\usepackage{amssymb}

\usepackage{setspace}
\usepackage{amsmath}

\newtheorem{definition}{Definition}

\newtheorem{example}{Example}

\begin{document}

\title{Modulation Diversity for Spatial Modulation Using Complex Interleaved Orthogonal Design}
\author{
      \IEEEauthorblockN{Rakshith Rajashekar and K.V.S. Hari} 
        \IEEEauthorblockA{Department of Electrical Communication Engineering\\ 
      Indian Institute of Science, Bangalore 560012\\ 
      \{rakshithmr, hari\}@ece.iisc.ernet.in}

\thanks{The financial support of the DST, India under the auspices of the India-UK Advanced Technology Centre (IU-ATC) is gratefully acknowledged.}

}
\maketitle

\begin{abstract}

In this paper, we propose modulation diversity techniques for Spatial Modulation (SM) system using Complex Interleaved Orthogonal Design (CIOD) meant for two transmit antennas. Specifically, we show that by using the CIOD for two transmit antenna system, the standard SM scheme, where only one transmit antenna is activated in any symbol duration, can achieve a transmit diversity order of two. We show with our simulation results that the proposed schemes offer transmit diversity order of two, and hence, give a better Symbol Error Rate performance than the SM scheme with transmit diversity order of one. 

\end{abstract}


\section{Introduction}

Spatial Modulation (SM) \cite{SM1}, \cite{SM2} is a novel low-complexity Multiple-Input Multiple-Output (MIMO) scheme that exploits MIMO channel for transmitting information in an innovative fashion. Specifically, the information bitstream is divided into blocks of length $\log_2({N_tM})$ bits, and in each block, $\log_2{(M)}$ bits select a symbol $s$ from an $M$-ary signal set (such as $M$-QAM or -PSK), and $\log_2{(N_t)}$ bits select an antenna out of $N_t$ transmit antennas for the transmission of the symbol $s$. The throughput achieved by this scheme is $R_{SM}=\log_2{(N_tM)}$ bpcu. Thus, the SM scheme achieves an increase in the spectral efficiency of $\log_2{N_t}$ bits over single-antenna systems with a marginal increase in the complexity as it still needs only one RF chain at the transmitter. One of the important benefits of the this scheme is the complete removal of Inter-Antenna Element interference at the receiver due to the activation of only one transmit antenna during any symbol duration. Since, only one transmit antenna is active at any instant the achievable transmit diversity will be limited to one. Some recent developments to increase the diversity order in SM scheme beyond one are Coherent Space-Time Shift Keying (CSTSK) \cite{CSTSK}, Time-Orthogonal Signal Design assisted Spatial Modulation (TOSD-SM) \cite{TOSD-SM}, \cite{TOSD-SSK}, and Space-Time Block Coded Spatial Modulation (STBC-SM) \cite{STBC-SM} schemes. 

The CSTSK scheme needs activation of all the transmit antennas in any symbol duration, and hence, needs $N_t$ RF chains at the transmitter. The STBC-SM scheme needs two active antennas out of $N_t$ over two channel uses for the transmission of the Alamouti code, thus needs two RF chains. In the TOSD-SM scheme, the diversity order of two is achieved by exploiting the independent delays associated with the channel gains, and needs only one RF chain. But, this scheme needs a relatively higher bandwidth due to the employment of non-sinusoid time-orthogonal signals. Table I summarizes the various properties of the existing schemes discussed above.

\begin{table*}
\begin{center}
\caption{Comparison of the existing schemes.}
\begin{tabular}{|c||c|c|c|c|}\hline
Properties & SM & CSTSK & TOSD-SM & STBC-SM\\
\hline\hline
$N_a$ & 1 & $N_t$ & 1 & 2\\
\hline
    IAS Required ? & No & Yes & No & Yes\\
\hline
  Tx Diversity order & 1 & $N_t$ & 2 & 2\\
\hline
\end{tabular}
\end{center}
\hspace{123pt}{$N_a:=$ no. of active antennas in a symbol duration.}\\
\hspace{120pt}{$~~~~~~~~~~~~~~~~~~~~~~~~~~~~~~~~~~~$IAS $:=$ Inter-Antenna Synchronization.}\\
\hrule
\end{table*}

The focus of this paper is to increase the diversity order in the SM system through coding over transmitted symbols without incurring rate loss or system complexity. We show that the Complex Interleaved Orthogonal Design (CIOD) \cite{SSDCIOD} meant for two transmit antennas can be employed in the SM system to achieve a transmit diversity order of two, while still activating only one transmit antenna in any symbol duration.

\begin{definition}
 The number of jointly used antenna combinations to convey information in an SM system is defined as the Degree of Spatial Modulation (DoSM). 
\end{definition}

For example, in the standard SM system the activation of one of the $N_t$ antennas conveys $\log_2{N_t}$ bits per channel use. Therefore, its DoSM is $N_t$. In the STBC-SM scheme, the throughput achieved is given by 
\[ R_{STBC-SM}=\frac{\log_2c + \log_2 M^2}{2} ~\text{bpcu}, \]
where, $M^2$ is the total number of combinations of two transmitted symbols in the Alamouti code, each taking values from a signal set of size $M$, and $c$ is the number of possible antenna combinations for the transmission of the Alamouti code, given by $c=\lfloor{ N_t \choose 2} \rfloor_{2^p}=\lfloor{ \frac{N_t(N_t-1)}{2}} \rfloor_{2^p}$, where $p$ is a positive integer. This scheme has the DoSM $c$. 

We note here that the maximum achievable DoSM per channel use in a system with $N_t$ transmit antennas is $N_t$. The effective DoSM over two channel uses in the standard SM scheme is $N_t^2$. It is easy to see that in the STBC-SM scheme $c< N_t^2$ for any $p$. In this paper, we propose two coding schemes, one with the DoSM $N_t$ over two channel uses and the other with the DoSM $N_t$ per channel use.

The rest of the paper is organized as follows. Section II briefly describes the SM system model. 
In Section III, the CIOD for two transmit antennas is reviewed. Sections IV and V respectively discuss the proposed schemes for the SM system to achieve a transmit diversity order of two with the DoSM $N_t$ over two channel uses and with the DoSM $N_t$ per channel use. Section VI discusses the ML decoding complexities of the proposed schemes. Simulation results and discussions are presented in Section VII. Section VIII concludes the paper.

\section{System model}

We consider a MIMO system having $N_t$ transmit as well as $N_r$ receive antennas and a quasi-static, frequency-flat fading channel, yielding:
\begin{equation}
 \mathbf{y}=\mathbf{H}\mathbf{x}+\mathbf{n},
\label{MIMOMODEL}
\end{equation}
where $\mathbf{x} \in \mathbb C^{N_t \times 1}$ is the transmitted vector, $\mathbf{y} \in \mathbb C^{N_r \times 1}$ is the received vector, $\mathbf{H} \in \mathbb C^{N_r \times N_t}$ is the channel matrix, and $\mathbf{n} \in \mathbb C^{N_r \times 1}$ is the noise vector. The entries of the channel matrix and the noise vector are from circularly symmetric complex-valued Gaussian distributions $\cal{CN}$(0,1) and $\cal{CN}$ (0, $2\sigma^2$), respectively, where ${\sigma^2}$ is the noise variance per dimension.

\subsection{Spatial Modulation}
In SM scheme \cite{SM1}, we have
\begin{equation}
 \mathbf{x}=[\underbrace{0,\hdots,0}_{l-1},s,\underbrace{0,\hdots,0}_{N_t-l}]^T \in \mathbb{C}^{N_t\times 1},
\label{SMTXVEC}
\end{equation}
where $s$ is a complex symbol from the signal set $S$ with $|S|=M$. Throughout this paper we assume $S$ to be a lattice constellation such as QAM. Thus, for an SM system, Eq.(\ref{MIMOMODEL}) becomes
\begin{equation}
 \mathbf{y}=\mathbf{H}\mathbf{x}_{l,s}+\mathbf{n},
\label{SMMODEL}
\end{equation}
where $l \in L=\{i\}_{i=1}^{N_t}$ and the subscript $s$ captures the dependence of $\mathbf x$ on the signal set $S$. Assuming perfect Channel State Information (CSI) and Maximum Likelihood (ML) decoding at the receiver, we have
\begin{align}
\label{MLSOLN}
(\hat{l},\hat{s})_{ML} &=\arg \min_{l\in L,s\in S} \| \mathbf{y}-\mathbf{H}\mathbf{x}_{l,s} \|_{2}^2,\\
 &=\arg \min_{l\in L,s\in S} \left\lbrace \sum_{i=1}^{N_r}|y_i-h_{l,i}s|^2 \right\rbrace,
\end{align}
where $y_i$ and $h_{l,i}$ are the $i^{\text{th}}$ and the $(l,i)^{\text{th}}$ entry of the received vector $\mathbf y$ and the channel matrix $\mathbf{H}$, respectively. 

\section{Review of CIOD for two transmit antennas }

The CIODs \cite{SSDCIOD} are basically symbol-by-symbol decodable codes which offer full rate (one complex symbol per channel use) for transmit antennas up to four. 

\subsection{CIOD for two transmit antennas}

The CIOD for two transmit antennas is given by
\begin{equation}
{\scriptsize \text{antennas}} {\Big\{} \underbrace{ \left[ \begin{array}{cc}
  \tilde s_1 & 0\\
  0 & \tilde s_2
 \end{array} \right] }_{\text{channel uses}}, \label{CIOD2TX}
\end{equation}
where, $\tilde s_1=s_{1I}+js_{2Q}$ and $\tilde s_2=s_{2I}+js_{1Q}$ are the transmitted symbols which are obtained by swapping the imaginary part of  $ s_1$ with that of $s_2$. We see from Eq.(\ref{CIOD2TX}) that in the first channel use $\tilde s_1$ is transmitted through the first antenna and the second antenna is kept inactive, and in the second channel use $\tilde s_2$ is transmitted through the second antenna and the first antenna is kept inactive. Thus, it is straightforward that this code uses only one active antenna during any channel use.

%

\subsection{Coding gain of the full rank STBC from CIOD}

Throughout this paper, we assume the signal set to be a square-QAM, represented by $S=\{d(2k-1-\sqrt M)+jd(2k-1-\sqrt M)~ | ~k,l \in [1,\sqrt M]\}$ with $|S|=M$ and $M$ is a perfect square, and $d$ is a scaling factor to ensure that the signal set has unit average energy.
\begin{definition}
 The Coordinate Product Distance (CPD) between any two signal points $x=x_I+jx_Q$ and $y=y_I+jy_Q$, $x\neq y$ in a signal set $S$ is defined as 
\begin{equation}
 CPD(x,y)=|x_I-y_I||x_Q-y_Q|,
\end{equation}
and the CPD of the signal set $S$ is defined as $CPD(S)=\min_{x,y \in S,x\neq y} CPD(x,y)$. 
\end{definition}

It was shown in \cite{SSDCIOD} that the full rank Space-Time Block Code (STBC) from CIOD gives full diversity when the signal set over which the STBC is constructed has non-zero CPD, and its coding gain was shown to be equal to the CPD of the signal set. Further, the coding gain of the STBC from CIOD constructed over square-QAM signal sets was shown to be maximized by rotating the constellation counterclockwise by $\theta=\frac{\arctan{(2)}}{2}$ radians, and the corresponding maximum coding gain achieved is $\frac{4d^2}{\sqrt 5}$. Thus, the STBC obtained from the design given in Eq.(\ref{CIOD2TX}) will have full diversity and maximum coding gain when the symbols $s_i=e^{j\theta}x_i$, where $x_i\in S$ for $i=\{1,2\}$.

\section{Proposed transmit diversity scheme with DoSM $N_t$ over two channel uses}
Considering the SM system given in Eq.(\ref{SMMODEL}) over two channel uses, we have
\begin{equation}
 [ \mathbf{y}_{1} ~{\mathbf{y}}_{2}]=\mathbf{H}\underbrace{[ {\mathbf{x}_{1}} ~{\mathbf{x}_{2}}]}_{\mathbf X}+ [{\mathbf{n}}_{1} ~{\mathbf{n}}_{2}],
\label{CIODSMMODEL1}
\end{equation}
where the subscript $l,s$ of the transmitted vector is dropped for the ease of presentation, the subscript $(\cdot)_{i}$ for $i=1,2$ indicates the channel use index, and the channel matrix $\mathbf H$ is assumed to be constant over two channel uses. Now, consider the transmission of the STBC obtained from two antenna CIOD given in Eq.(\ref{CIOD2TX}) in the above SM system. The first symbol ${\tilde s}_1$ can be transmitted through one of the $N_t$ antennas in ${\mathbf x}_1$.  Note that the second symbol ${\tilde s}_2$ cannot be transmitted through the same antenna that was used in the transmission of ${\tilde s}_1$ in order to achieve full diversity, so, ${\tilde s}_2$ can be transmitted through one of the $N_t-1$ antennas in ${\mathbf x}_2$. To have DoSM $N_t$ over two channel uses we choose one of the $N_t$ Code Books (CB) given in Eq.(\ref{NTDOSMCB}). Observe from Eq.(\ref{NTDOSMCB}) that if the symbol ${\tilde s}_1$ is transmitted through the antenna $l$ in the first channel use then the symbol ${\tilde s}_2$ is transmitted through the antenna $(l+1)\mod N_t$ in the second channel use. The rate achieved by this scheme is

\begin{figure*}
\centering
\begin{equation}
\left\{ \underbrace{\left[ \begin{array}{cc}
{\tilde s}_1 & 0\\
0 & {\tilde s}_2\\
0 & 0\\
\vdots & \vdots\\
0 & 0\\
0 & 0
\end{array} \right]}_{\text{CB}_1},
 \underbrace{\left[ \begin{array}{cc}
0 & 0\\
{\tilde s}_1 & 0\\
0 & {\tilde s}_2\\
\vdots & \vdots\\
0 & 0\\
0 & 0
\end{array} \right]}_{\text{CB}_2},
\underbrace{\left[ \begin{array}{cc}
0 & 0\\
0 & 0\\
{\tilde s}_1 & 0\\
0 & {\tilde s}_2\\
\vdots & \vdots\\
0 & 0
\end{array} \right]}_{\text{CB}_3},\hdots\hdots\hdots,
 \underbrace{\left[ \begin{array}{cc}
0 & 0\\
0 & 0\\
0 & 0\\
\vdots & \vdots\\
{\tilde s}_1 & 0\\
0 & {\tilde s}_2
\end{array} \right]}_{\text{CB}_{N_t-1}},
 \underbrace{\left[ \begin{array}{cc}
0 & {\tilde s}_2\\
0 & 0\\
0 & 0\\
\vdots & \vdots\\
0 & 0\\
{\tilde s}_1 & 0
\end{array} \right]}_{\text{CB}_{N_t}} \right\}
\label{NTDOSMCB}
\end{equation}
\hrule
\end{figure*}

\begin{equation}
 \frac{\log_2N_t + \log_2 M^2}{2}~\text{bpcu}.
\end{equation}

\subsection{Diversity and Coding gain}

Let the coding gain of the proposed scheme be defined by
\begin{equation}
G=\min_{{\mathbf X}\neq{\mathbf X}^\prime} |\det{\Delta^H\Delta}|,
\end{equation}
where, $\Delta={\mathbf X}-{\mathbf X}^\prime$, and ${\mathbf X}$, ${\mathbf X}^\prime$ $\in {\mathbb C}^{N_t\times2}$ are the transmitted space-time matrices from the same or different Code Books. When they belong to the same Code Book, say $\text{CB}_i$ for any $1\leq i\leq N_t$, we have
\begin{equation}
 \Delta^H\Delta=\left[ \begin{array}{cc}
|{\tilde s}_1-{\tilde s}_1^\prime|^2 & 0\\
0 & |{\tilde s}_2-{\tilde s}_2^\prime|^2
\end{array} \right].
\end{equation}
Thus we have, 
\begin{equation}
G=\min_{{{\tilde s}_1,{\tilde s}_2,{\tilde s}_1^\prime,{\tilde s}_1^\prime},{\tilde s}_1\neq {\tilde s}_1^\prime} |{\tilde s}_1-{\tilde s}_1^\prime|^2 |{\tilde s}_2-{\tilde s}_2^\prime|^2 
\end{equation}
which is nothing but the coding gain of the CIOD for two transmit antennas. Note that the symbols ${\tilde s}_1$ and ${\tilde s}_2$ in each of the Code Books are obtained by interleaving the imaginary parts of $s_1$ and $s_2$, where $s_i=e^{j\theta}x_i$, $x_i\in S$ for $i=\{1,2\}$, and $\theta=\frac{\arctan{(2)}}{2}$. This ensures that when ${\mathbf X}$, ${\mathbf X}^\prime$ $\in \text{CB}_i$, the ${rank}(\Delta^H\Delta)=2$, and $G$ is maximized. 

Now consider the case when ${\mathbf X}\in \text{CB}_i$ and ${\mathbf X}^\prime\in \text{CB}_j$ where $j=(i+1)\mod N_t$ or $(i-1)\mod N_t$. Without loss of generality we have
\begin{equation}
 \Delta^H\Delta=\left[ \begin{array}{cc}
|{\tilde s}_1|^2+|{\tilde s}_1^\prime|^2 & {-{\tilde s}_1^{\prime *}{\tilde s}_2}\\
{-{\tilde s}_1^\prime{\tilde s}_2^*} & |{\tilde s}_2|^2+|{\tilde s}_2^\prime|^2
\end{array} \right].
\end{equation}
Thus, we have 
\begin{align}
\det(\Delta^H\Delta) &=(|{\tilde s}_1|^2+|{\tilde s}_1^\prime|^2) (|{\tilde s}_2|^2+|{\tilde s}_2^\prime|^2)-{|{\tilde s}_1^\prime|^2|{\tilde s}_2|^2},\\
&={|{\tilde s}_1|^2|{\tilde s}_2|^2}+{|{\tilde s}_1|^2|{\tilde s}_2^\prime|^2}+{|{\tilde s}_1^\prime|^2|{\tilde s}_2^\prime|^2}.
\end{align}
It is easy to see that the $\det(\Delta^H\Delta)\neq 0$ for any $s_i,s_i^\prime\in e^{j\theta}S$.

Now consider the case when ${\mathbf X}\in \text{CB}_i$ and ${\mathbf X}^\prime\in \text{CB}_j$ where $j\neq i$ or $(i+1)\mod N_t$ or $(i-1)\mod N_t$. In this case, without loss of generality, we have
\begin{equation}
\det(\Delta^H\Delta) =(|{\tilde s}_1|^2+|{\tilde s}_1^\prime|^2) (|{\tilde s}_2|^2+|{\tilde s}_2^\prime|^2).
\end{equation}
It is straightforward that the $\det(\Delta^H\Delta)\neq 0$ for any $s_i,s_i^\prime\in e^{j\theta}S$. Thus, the matrices from the collection of Code Books given in Eq.(\ref{NTDOSMCB}) offer diversity order of two.

In the rest of this paper, this scheme is referred to as the {\em CIOD-SM scheme with Low DoSM}.

\section{Proposed transmit diversity scheme with DoSM $N_t$ per channel use}
In the previous section we proposed a CIOD based SM scheme with DoSM $N_t$ over two channel uses. In this section we propose a CIOD based SM scheme with DoSM $N_t$ per channel use. Suppose, if we have $N_t+1$ transmit antennas then we could transmit ${\tilde s}_1$ through one of the first $N_t$ antennas and ${\tilde s}_2$ through one of the $N_t$ antennas  excluding the antenna used in the transmission of ${\tilde s}_1$.
That is,
\begin{equation}
 [ {\mathbf{x}_{1}} ~{\mathbf{x}_{2}}]=\left[  \begin{array}{cc}
						0 & 0\\
						0 & {\tilde s}_2\\
						\vdots & \vdots\\
						0 & 0\\
						{\tilde s}_1 & *\\
						0 & 0\\
						\vdots & \vdots\\
						0 & 0\\
						\bullet & 0
                                               \end{array}
 \right]_{(N_t+1)\times 2},
\end{equation}
where, the $\bullet$ in the $(N_t+1)^{\text{th}}$ row of the first column indicates that the $(N_t+1)^{\text{th}}$ antenna is never used in the transmission of ${\tilde s}_1$, where as the $*$ in the second column adjacent to ${\tilde s}_1$ indicates that the antenna used for the transmission of ${\tilde s}_1$ will not be used in the transmission of ${\tilde s}_2$. If $s_1, s_2 \in e^{j\theta}S$, where $S$ is a square $M$-QAM, then $s_1$ and $s_2$ together can take $M^2$ values. Also, from the above discussion we have $N_t^2$ number of antenna activations possible in two channel uses. Thus, the rate achieved by this scheme is
\begin{equation}
 \frac{\log_2N_t^2 + \log_2 M^2}{2}=\log_2{(N_tM)}~\text{bpcu},
\end{equation}
which is same as that of the standard SM scheme. Note that there are $N_t^2$ number of Code Books in this scheme.

\subsection{Diversity and Coding gain}

The following example illustrates the method of obtaining the Code Books for SM scheme with DoSM $N_t$ per channel use that give a transmit diversity order of two.

\begin{example}
 
Consider an SM system with the number of transmit antennas $N_t+1=5$ and the proposed CIOD based SM scheme with DoSM $N_t$ per channel use discussed in the previous subsection. The number of Code Books in this scheme is 16. Consider the following two Code Books.

\begin{equation}
 \left[  \begin{array}{cc}                      {\tilde s}_1 & *\\
						0 & {\tilde s}_2\\
						0 & 0\\
						0 & 0\\
						\bullet & 0
                                               \end{array} \right],
\left[  \begin{array}{cc}                       {\tilde s}_1 & *\\
						0 & 0\\
						0 & {\tilde s}_2\\
						0 & 0\\
						\bullet & 0
                                               \end{array} \right].
\end{equation}
It is easy to see that the difference of the matrices from these Code Books will have rank 1 when ${\tilde s}_1$ in the first Code Book above is same as that of the second. We refer to such Code Books as interfering Code Books where in the difference of the matrices chosen from each of them would be rank deficient. The 16 Code Books are grouped into four Code Book Sets as shown below. It can be verified that the Code Books within a Code Book Set (CBS) do not interfere.

\begin{equation}
\footnotesize
CBS_1= \left\{  \left[  \begin{array}{cc}       {\tilde s}_1 & *\\
						0 & {\tilde s}_2\\
						0 & 0\\
						0 & 0\\
						\bullet & 0
                                               \end{array} \right],
\left[  \begin{array}{cc}                       0 & 0\\
						{\tilde s}_1 & *\\
						0 & {\tilde s}_2\\
						0 & 0\\
						\bullet & 0
                                               \end{array} \right],
 \left[  \begin{array}{cc}                      0 & 0\\
						0 & 0\\
						{\tilde s}_1 & *\\
						0 & {\tilde s}_2\\
						\bullet & 0
                                               \end{array} \right],
\left[  \begin{array}{cc}                       0 & 0\\
						0 & 0\\
						0 & 0\\
						{\tilde s}_1 & *\\
						\bullet & {\tilde s}_2
                                               \end{array} \right] \right\}
\end{equation}
\begin{equation}
\footnotesize
CBS_2= \left\{  \left[  \begin{array}{cc}       {\tilde s}_1 & *\\
						0 & 0\\
						0 & {\tilde s}_2\\
						0 & 0\\
						\bullet & 0
                                               \end{array} \right],
\left[  \begin{array}{cc}                       0 & 0\\
						{\tilde s}_1 & *\\
						0 & 0\\
						0 & {\tilde s}_2\\
						\bullet & 0
                                               \end{array} \right],
 \left[  \begin{array}{cc}                      0 & 0\\
						0 & 0\\
						{\tilde s}_1 & *\\
						0 & 0\\
						\bullet & {\tilde s}_2
                                               \end{array} \right],
\left[  \begin{array}{cc}                       0 & {\tilde s}_2\\
						0 & 0\\
						0 & 0\\
						{\tilde s}_1 & *\\
						\bullet & 0
                                               \end{array} \right] \right\}
\end{equation}
\begin{equation}
\footnotesize
CBS_3= \left\{  \left[  \begin{array}{cc}       {\tilde s}_1 & *\\
						0 & 0\\
						0 & 0\\
						0 & {\tilde s}_2\\
						\bullet & 0
                                               \end{array} \right],
\left[  \begin{array}{cc}                       0 & 0\\
						{\tilde s}_1 & *\\
						0 & 0\\
						0 & 0\\
						\bullet & {\tilde s}_2
                                               \end{array} \right],
 \left[  \begin{array}{cc}                      0 & {\tilde s}_2\\
						0 & 0\\
						{\tilde s}_1 & *\\
						0 & 0\\
						\bullet & 0
                                               \end{array} \right],
\left[  \begin{array}{cc}                       0 & 0\\
						0 & {\tilde s}_2\\
						0 & 0\\
						{\tilde s}_1 & *\\
						\bullet & 0
                                               \end{array} \right] \right\}
\end{equation}
\begin{equation}
\footnotesize
CBS_4= \left\{  \left[  \begin{array}{cc}       {\tilde s}_1 & *\\
						0 & 0\\
						0 & 0\\
						0 & 0\\
						\bullet & {\tilde s}_2
                                               \end{array} \right],
\left[  \begin{array}{cc}                       0 & {\tilde s}_2\\
						{\tilde s}_1 & *\\
						0 & 0\\
						0 & 0\\
						\bullet & 0
                                               \end{array} \right],
 \left[  \begin{array}{cc}                      0 & 0\\
						0 & {\tilde s}_2\\
						{\tilde s}_1 & *\\
						0 & 0\\
						\bullet & 0
                                               \end{array} \right],
\left[  \begin{array}{cc}                       0 & 0\\
						0 & 0\\
						0 & {\tilde s}_2\\
						{\tilde s}_1 & *\\
						\bullet & 0
                                               \end{array} \right] \right\}
\end{equation}

In order to achieve full diversity, each of the CBSs given above is multiplied by a complex exponential and the phase angles are obtained through numerical simulations by maximizing the coding gain of the effective Code Book given by $\{e^{j\theta_1}CBS_1,~e^{j\theta_2}CBS_2,~e^{j\theta_3}CBS_3,~e^{j\theta_4}CBS_4\}$.
$~~~~~~~~~~~~~\blacksquare$
\end{example}

In general, in an SM system with $N_t+1$ number of transmit antennas, where $N_t$ is a power of two, the $N_t$ CBSs each containing $N_t$ CBs are given by 
\begin{equation}
 CBS_i=\{e^{j\theta_i}CB_{i,j} ~|~ 1\leq j\leq N_t\} ~\text{for}~ 1\leq i\leq N_t,
\end{equation}
where, $CB_{i,j}$ is an $(N_t+1)\times 2$ matrix with two non zero elements ${\tilde s}_1$ and ${\tilde s}_2$ at $(j,1)$ and $((j+i)\mod(N_t+1),2)$, respectively.

In the rest of this paper, this scheme is referred to as the {\em CIOD-SM scheme with High DoSM}.

\section{ML Decoding complexity of the proposed schemes}

In this section we discuss the ML decoding of the proposed schemes with their complexities.
Eq.(\ref{CIODSMMODEL1}) can written as
\begin{equation}
  \mathbf{y}_{1} ={\mathbf{h}}_{l_1} {\tilde{s}_{1}} + {\mathbf{n}}_{1},
\label{CIODSM1}
\end{equation}
\begin{equation}
  \mathbf{y}_{2} ={\mathbf{h}}_{l_2} {\tilde{s}_{2}} + {\mathbf{n}}_{2},
\label{CIODSM2}
\end{equation}
where, $1\leq l_1 \leq N_t$ and $l_2=(l_1+1)\mod N_t$ in the proposed CIOD-SM scheme with Low DoSM, and $1\leq l_1 \leq N_t$, $1\leq l_2 \leq N_t+1, ~l_2\neq l_1$ in the proposed CIOD-SM scheme with High DoSM. For the ease of presentation, the exponentials associated with the different CBSs of the proposed High DoSM scheme is absorbed in ${\mathbf{h}}_{l_1}$ and ${\mathbf{h}}_{l_2}$. The ML solution can be written as 
\begin{equation}
\small
({\hat{l}}_1,{\hat{l}}_2,\hat{s}_1,\hat{s}_2 )_{ML} =
\arg \min_{\text{legitimate}~l_1,l_2 } \left[ \min_{s_1,s_2\in \{ e^{j\theta}S\}} J(l_1,l_2,s_1,s_2)\right],\\
\label{2CHMLSOLN}
\end{equation}
where $J(l_1,l_2,s_1,s_2)=\| {\mathbf{y}}_1-{\mathbf{h}}_{l_1}{\tilde s}_{1} \|_{2}^2 + \| {\mathbf{y}}_2-{\mathbf{h}}_{l_2}{\tilde s}_{2} \|_{2}^2$.
For given ${\mathbf h}_{l_1}$ and ${\mathbf h}_{l_2}$, Maximal Ratio Combining (MRC) yields
\begin{equation}
 {\hat y}_1 =a {\tilde{s}_{1}} + {\hat n}_1,
\label{CIODSMMRC1}
\end{equation}
\begin{equation}
   {\hat y}_2 =b {\tilde{s}_{2}} + {\hat n}_2 ,
\label{CIODSMMRC2}
\end{equation}
where, $ {\hat y}_1={\mathbf{h}}_{l_1}^H  \mathbf{y}_{1}$, ${\hat y}_2 = {\mathbf{h}}_{l_2}^H\mathbf{y}_{2}$, $a=\| {\mathbf{h}}_{l_1}\|^2$, $b=\| {\mathbf{h}}_{l_2}\|^2$, ${\hat n}_1={\mathbf{h}}_{l_1}^H{\mathbf{n}}_{1}$, and ${\hat n}_2= {\mathbf{h}}_{l_2}^H {\mathbf{n}}_{2}$. Note that 
${\hat n}_1 \sim {\cal CN}(0,2a\sigma^2)$ and ${\hat n}_2 \sim {\cal CN}(0,2b\sigma^2)$. Equations (\ref{CIODSMMRC1}) and (\ref{CIODSMMRC2}) can be written as
\begin{equation}
 {\hat y}_1 =a {(s_{1I}+js_{2Q})} + {\hat n}_1,
\end{equation}
\begin{equation}
   {\hat y}_2 =b {(s_{2I}+js_{1Q})} + {\hat n}_2.
\end{equation}
Upon de-interleaving the imaginary parts we get,
\begin{equation}
 {\tilde y}_1 =a s_{1I}+j bs_{1Q} + {\tilde n}_1,
\label{CIODSMDIV1}
\end{equation}
\begin{equation}
   {\tilde y}_2 =b s_{2I}+j as_{2Q} + {\tilde n}_2,
\label{CIODSMDIV2}
\end{equation}
where, ${\tilde y}_1={\hat y}_{1I}+j{\hat y}_{2Q}$, ${\tilde y}_2={\hat y}_{2I}+j{\hat y}_{1Q}$, ${\tilde n}_1={\hat n}_{1I}+j{\hat n}_{2Q}$, ${\tilde n}_2={\hat n}_{2I}+j{\hat n}_{1Q}$. Note that ${\hat n}_{1I} \sim {\cal N}(0,a\sigma^2)$, ${\hat n}_{2Q} \sim {\cal N}(0,b\sigma^2)$, and ${\hat n}_{2I} \sim {\cal N}(0,b\sigma^2)$, ${\hat n}_{1Q} \sim {\cal N}(0,a\sigma^2)$, thus, the real and imaginary parts of ${\tilde n}_{1}$ and ${\tilde n}_{2}$ have different variances. The ML solution \cite{SSDCIOD} in this case is given as follows:

Choose $s_1=s_i \in e^{j\theta}S$ such that 
\begin{equation}
 b|{\hat y}_1-as_{iI}|^2 +a|{\hat y}_1-bs_{iQ}|^2 \leq b|{\hat y}_1-as_{kI}|^2 +a|{\hat y}_1-bs_{kQ}|^2 
\label{MLCIODSOLN1}
\end{equation}
$\forall ~i\neq k$.

Choose $s_2=s_i\in e^{j\theta}S$ such that 
\begin{equation}
 a|{\hat y}_2-bs_{iI}|^2 +b|{\hat y}_2-as_{iQ}|^2 \leq a|{\hat y}_2-bs_{kI}|^2 +b|{\hat y}_2-as_{kQ}|^2 
\label{MLCIODSOLN2}
\end{equation}
$\forall ~i\neq k$.

Thus, from Eq.(\ref{CIODSMDIV1}) and Eq.(\ref{CIODSMDIV2}) it is clear that both $s_1$ and $s_2$ enjoy the diversity order of at least two (when $N_r=1$) due to the independent fading coefficients associated with their real and imaginary parts, and from Eq.(\ref{MLCIODSOLN1}) and Eq.(\ref{MLCIODSOLN2}) it is clear that the the symbols $s_1$ and $s_2$ can be decoded independently resulting in a complexity of $2M$ instead of $M^2$. Also, notice that the transmit diversity order reduces to one if the symbols $s_1$ and $s_2$ are from the unrotated square-QAM signal set, and by rotating the constellation, the real and imaginary parts are coded and hence the diversity order increases to two.

For a given $(l_1,l_2)$ pair, $s_1$ and $s_2$ are obtained as above. Since the number of combinations of $(l_1,l_2)$ is $N_t$ for the proposed scheme with Low DoSM, the total number of evaluations of Eq.(\ref{MLCIODSOLN1}) and Eq.(\ref{MLCIODSOLN2}) in obtaining the ML solution of Eq.(\ref{2CHMLSOLN}) will be $2MN_t$. Note that the SM scheme employing a square $M$-QAM signal set with $N_t$ transmit antennas will have the search complexity of $2{M}N_t$ over two channel uses which is same as that of the proposed scheme with Low DoSM. The number of combinations of $(l_1,l_2)$ for the proposed scheme with High DoSM is $N_t^2$. Hence, the search complexity of this scheme is $2MN_t^2$ which is $N_t\times 2{M}N_t$. Thus, the proposed scheme with High DoSM has an order $N_t$ higher search complexity than the standard SM scheme. However, we show in the next section that the search complexities of the proposed schemes can be further reduced by exploiting the structure of the CIOD.

\section{Low complexity ML detector for our proposed schemes}

In the previous section, it was show that the ML decoding complexity of our proposed scheme with Low DoSM is $2N_tM$ and that of the High DoSM scheme is $2N_t^2M$. In this section, we show that the ML decoding complexities of the proposed schemes can be further reduced. Specifically, we show that the ML decoding complexity in case of Low DoSM scheme can be reduced to $2N_t\sqrt{M}$ and that in case of High DoSM scheme can be reduced to $2N_t^2\sqrt{M}$.

Equations (\ref{CIODSM1}) and (\ref{CIODSM2}) can be written as
\begin{equation}
\left[ \mathbf{y}_{1} ~\mathbf{y}_{2}\right] = \left[ {\mathbf{h}}_{l_1} ~{\mathbf{h}}_{l_2}\right] \left[\begin{array}{cc}
                                                                                                      {\tilde{s}_{1}} & 0\\
												      0 & {\tilde{s}_{2}}
                                                                                                     \end{array}\right] +\left[{\mathbf{n}}_{1} ~{\mathbf{n}}_{2} \right]. 
\end{equation}
Upon vectorizing the above equation we arrive at
\begin{equation}
 \left[ \begin{array}{c}
         {\mathbf y}_1\\
	 {\mathbf y}_2
        \end{array}
\right]=\left[ {\mathbf I}_2 \otimes {\mathbf H}(l_1,l_2) \right] 
\left[ \begin{array}{c}
               {\tilde{s}_{1}} \\
	      0 \\
	      0 \\
	      {\tilde{s}_{2}} 
              \end{array}
\right]
+\left[ \begin{array}{c}
                {\mathbf n}_1\\
		{\mathbf n}_2
               \end{array}
\right],\label{CIODSM3}
\end{equation}
where ${\mathbf I}_2$ is a $2\times 2$ identity matrix, ${\mathbf H}(l_1,l_2)=\left[ {\mathbf{h}}_{l_1} ~{\mathbf{h}}_{l_2}\right]$, and $\otimes$ represents the Kronecker product. Also, we may write
\begin{align}
\small
 \left[ \begin{array}{c}
               {\tilde{s}_{1}} \\
	      0 \\
	      0 \\
	      {\tilde{s}_{2}} 
              \end{array}
\right]&=
\underbrace{\left[ \begin{array}{cccc}
                                                                 1 & 0 & 0 & j\\
								 0 & 0 & 0 & 0\\
								 0 & 0 & 0 & 0\\
								 0 & j & 1 & 0
       \end{array}
\right]}_{{\mathbf V}_1}\left[ \begin{array}{c}
               s_{1I} \\
	      s_{1Q} \\
	      s_{2I} \\
	      s_{2Q} 
              \end{array}
\right], \text{ and}\\
\small
\left[ \begin{array}{c}
               s_{1I} \\
	      s_{1Q} \\
	      s_{2I} \\
	      s_{2Q} 
              \end{array}
\right]&=
\small
\underbrace{\left[ \begin{array}{cccc}
                                                                 \cos(\theta) & -\sin(\theta) & 0 & 0\\
								 \sin(\theta) & ~~\cos(\theta) & 0 & 0\\
								 0 & 0 & \cos(\theta) & -\sin(\theta)\\
								 0 & 0 & \sin(\theta) & ~~\cos(\theta)
                                                                \end{array}
\right]}_{{\mathbf V}_2}
\small
\left[\begin{array}{c}
               x_{1I} \\
	      x_{1Q} \\
	      x_{2I} \\
	      x_{2Q} 
              \end{array}
\right].
\end{align}
Thus, Eq.(\ref{CIODSM3}) can be written as
\begin{equation}
 \left[ \begin{array}{c}
         {\mathbf y}_1\\
	 {\mathbf y}_2
        \end{array}
\right]=\underbrace{\left[ {\mathbf I}_2 \otimes {\mathbf H}(l_1,l_2) \right] {\mathbf{V}}_1 {\mathbf{V}}_2}_{{\mathbf H}_{eq}(l_1,l_2)}\left[ \begin{array}{c}
               x_{1I} \\
	      x_{1Q} \\
	      x_{2I} \\
	      x_{2Q} 
              \end{array}
\right]+\left[ \begin{array}{c}
                {\mathbf n}_1\\
		{\mathbf n}_2
               \end{array}
\right].
\end{equation}
The above equation can be expressed in terms of real variables{\footnote{Given a complex-valued vector ${\mathbf z}\in {\mathbb C}^{m}$ its equivalent real-valued vector is given by 
\begin{equation*}
 {\bar {\mathbf z}}=\left[ {z_{1I}}, {z_{1Q}},{z_{2I}}, {z_{2Q}},\hdots,{z_{mI}}, {z_{mQ}} \right] \in {\mathbb R}^{2m},
\end{equation*}
where $z_{iI}=\Re{(z_i)}$ and $z_{iQ}=\Im{(z_i)}$ for $1\leq i \leq m$.
Similarly, a complex valued matrix ${\mathbf A} \in {\mathbb C}^{m\times n}$ is represented in terms of real values by $\bar{\mathbf A} \in {\mathbb R}^{2m\times 2n}$ which is obtained by replacing every $(i,j)^{\text{th}}$ element of $\mathbf A$ by
\begin{equation*}
 \left[ \begin{array}{cc}
  \Re{(a_{i,j})} & -\Im{(a_{i,j})}\\
  \Im{(a_{i,j})} & ~~\Re{(a_{i,j})}
 \end{array} \right].
\end{equation*}}} as 
\begin{equation}
 \underbrace{\left[ \begin{array}{c}
         \bar{\mathbf y}_1\\
	 \bar{\mathbf y}_2
        \end{array}
\right]}_{\bar{\mathbf y}}={\bar{\mathbf H}_{eq}(l_1,l_2)}
\underbrace{\left[ \begin{array}{c}
               x_{1I} \\
	      x_{1Q} \\
	      x_{2I} \\
	      x_{2Q} 
              \end{array}
\right]}_{\bar{\mathbf x}}+\left[ \begin{array}{c}
             \bar   {\mathbf n}_1\\
		\bar {\mathbf n}_2
               \end{array}
\right].
\end{equation}
It can be verified that the columns ${\bar{\mathbf h}}_1$ and ${\bar{\mathbf h}}_2$ of $\bar{\mathbf H}_{eq}(l_1,l_2)=[{\bar{\mathbf h}}_1 ~{\bar{\mathbf h}}_2 ~{\bar{\mathbf h}}_3 ~{\bar{\mathbf h}}_4]$ are orthogonal to both the columns ${\bar{\mathbf h}}_3$ and ${\bar{\mathbf h}}_4$ for any realization of ${\mathbf H}(l_1,l_2)$.
 
The ML solution in terms of real-valued system is given by
\begin{equation} 
 ({\hat{l}}_1,{\hat{l}}_2,\hat{x}_{1I},\hat{x}_{1Q},\hat{x}_{2I},\hat{x}_{2Q} )_{ML} =
\end{equation}
\begin{equation}
 \arg \left\{ \min_{l_1,l_2 }\left[ \min_{{x}_{1I},{x}_{1Q},{x}_{2I},{x}_{2Q}} \|\bar {\mathbf y} - {\bar{\mathbf H}}_{eq}(l_1,l_2) \bar{\mathbf x} \|_2^2 \right]\right\},
\end{equation}
where ${x}_{1I},{x}_{2I}$ and ${x}_{1Q},{x}_{2Q}$ are assumed to take values from $N_1$-PAM and $N_2$-PAM signal sets respectively. For example, if $S$ is a 32-QAM signal set then it can be thought of as the Cartesian product of 8-PAM and 4-PAM signal sets along the real and imaginary axes, respectively. In case of a square $M$-QAM signal set the real and imaginary axes will have $\sqrt{M}$ PAM points each.
 
Taking the QR decomposition of ${\bar{\mathbf H}}_{eq}(l_1,l_2)$ we have 
\[\|\bar {\mathbf y} - {\bar{\mathbf H}}_{eq}(l_1,l_2) \bar{\mathbf x} \|_2^2=\|\bar {\mathbf y} - \bar{\mathbf Q} \bar{\mathbf R} \bar{\mathbf x} \|_2^2,\] where 
$\bar{\mathbf Q}=[\bar{\mathbf Q}_{1~4N_r\times 4} \bar{\mathbf Q}_{2~4N_r\times (4N_r-4)}]$ and $ \bar{\mathbf R}=\left[\begin{array}{c}
                                                                                \bar{\mathbf R}_{1~4\times4}\\
										   {\mathbf 0}_{(4N_r-4)\times4}
                                                                              \end{array}\right].
$
Thus, we have 
\[\|\bar {\mathbf y} - \bar{\mathbf Q} \bar{\mathbf R} \bar{\mathbf x} \|_2^2=\|\bar {\mathbf z}_1 - \bar{\mathbf R}_1 \bar{\mathbf x} \|_2^2+\|{\mathbf z}_2\|_2^2,\] where $\bar{\mathbf z}_1=\bar{\mathbf Q}_1^T\bar{\mathbf y}$ and $\bar{\mathbf z}_2=\bar{\mathbf Q}_2^T\bar{\mathbf y}$. Since, $\bar{\mathbf h}_1$ and $\bar{\mathbf h}_2$ are both orthogonal to $\bar{\mathbf h}_3$ and $\bar{\mathbf h}_4$, it is easy to see that the elements $r_{i,j}=0$ for $i=\{1,2\},j=\{3,4\}$, i.e, $\bar{\mathbf R}_1$ has the structure,
\begin{equation}
 \left[ \begin{array}{cccc}
        \star & \star & 0 & 0\\
	  0 & \star & 0 & 0\\
	  0 & 0 & \star & \star\\
	  0 & 0 & 0 & \star
       \end{array}\right]
\end{equation}
where, $\star$ indicates a non-zero element. It is clear that the symbols $s_1$ and $s_2$ can be decoded independently. Furthermore, upon conditioning $x_{1Q}$ and $x_{2Q}$, $x_{1I}$ and $x_{2I}$ can be decoded without searching through the signal set by {\em hard-limiting } as given by Eq.(\ref{HDL1}) and Eq.(\ref{HDL2}).
\begin{figure*}
\begin{equation}
\small
 \hat{x}_{1I}= \min \left[ \max \left( 2 ~rnd \left[ \frac{\left(\bar{z}_{1_1}-\bar{x}_{1Q}r_{1,2}\right)/r_{1,1}+1}{2}\right]-1,-\sqrt{N_1}+1\right),\sqrt{N_1}-1\right]
\label{HDL1}
\end{equation}
\end{figure*}
\begin{figure*}
\begin{equation}
\small
 \hat{x}_{2I}= \min \left[ \max \left( 2 ~rnd \left[ \frac{\left(\bar{z}_{1_3}-\bar{x}_{2Q}r_{3,4}\right)/r_{3,3}+1}{2}\right]-1,-\sqrt{N_2}+1\right),\sqrt{N_2}-1\right]\label{HDL2}
\end{equation}
\hrule
\end{figure*}
Assuming $N_1=N_2=\sqrt{M}$, we have $2\sqrt{M}$ as the complexity of conditioning $x_{1Q}$ and $x_{2Q}$. Thus, the proposed scheme with Low DoSM has the ML decoding complexity of $N_t(2\sqrt{M})$ and the proposed scheme with High DoSM has the ML decoding complexity of $N_t^2(2\sqrt{M})$.

\section{Simulation Results and Discussion}

Consider an SM system with $N_r=2$, $N_t=4$, employing BPSK, and 4-QAM signal sets. With the BPSK and the 4-QAM signal sets the spectral efficiencies achieved are 3 and 4 bpcu, respectively. The same spectral efficiencies are achieved by the proposed scheme with Low DoSM by employing 4-QAM and 8-QAM signal sets. The proposed scheme with High DoSM achieves the same spectral efficiencies by using the same signal sets as used by the SM scheme but needs five transmit antennas instead of four. In all our simulations, at an SER of $10^{-t}$ we have used at least $10^{t+1}$ symbols in evaluating the SER, and assumed block Rayleigh fading channel. The receiver is assumed to have perfect CSI and perform ML decoding.

Recall from Section IV that the coding gain $G$ of the proposed scheme with Low DoSM depends only on the transmitted symbols and not on the number of transmit antennas. Table II gives the coding gain offered by the proposed scheme with Low DoSM for various QAM signal sets. From Table II it is clear that the coding gain remains constant for $M\geq16$. {\em We conjecture that the proposed scheme with Low DoSM for an SM system with arbitrary $N_t$ has the Non Vanishing Determinant (NVD) property \cite{VITERBO}.}

\begin{table}
\caption{Coding gain offered by the proposed scheme with Low DoSM.}
\begin{center}
\begin{tabular}{|c||c|c|c|c|}\hline
  $M$-QAM &   $M=4$ &  $M=16$&  $M=64$ & $M=256$\\
	\hline\hline
  $G$  &   1.6446 &   1.6 &  1.6 &  1.6\\
\hline
\end{tabular}
\end{center}
\end{table}

The coding gain offered by the proposed scheme with High DoSM for BPSK and 4-QAM signal sets are 0.2 and 0.04, respectively. The complex exponential functions associated with the four CBSs in the proposed scheme with High DoSM are obtained by maximizing the coding gain over the sixteen points on the unit circle, and are given in Table III. Note that the higher coding gains than what is achieved may be obtained by maximizing the coding gain by considering a larger number of points on the unit circle.

\begin{table*}
\caption{Coding gain optimized exponentials for the proposed scheme with High DoSM.}
\begin{center}
\begin{tabular}{|c||c|c|c|c|}\hline
 \small  &   $e^{j\theta_1}$ &  $e^{j\theta_2}$&  $e^{j\theta_3}$ & $e^{j\theta_4}$\\
\hline
\hline
   BPSK&   $0.9239 + 0.3827j$ &   $0.7071-0.7071j$ &  $-1.0000j$ &  $-0.7071-0.707j$\\
\hline
   4-QAM  &   $-0.9239 + 0.3827j$ &   $1.0000j$ &  $0.7071 + 0.7071j$ &  $0.9239 + 0.3827j$\\
	\hline
\end{tabular}
\end{center}
\hrule
\end{table*}

\begin{figure}
\centering
\includegraphics[scale=0.36]{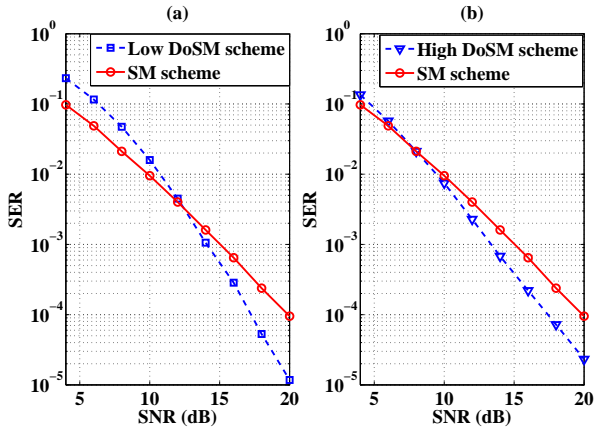}
\caption{Plot (a)  gives the SER curves of the proposed scheme with Low DoSM and the SM scheme. Plot (b) gives the SER curves of the proposed scheme with High DoSM and the SM scheme. In both the scenarios the spectral efficiency is 3 bpcu.}
\label{SERSM3BPCU}
\end{figure}

Fig. \ref{SERSM3BPCU} gives the Symbol Error Rate (SER) performance of the proposed schemes and the SM scheme for the spectral efficiency of 3 bpcu. It is clear from Fig. \ref{SERSM3BPCU} that both the proposed schemes exhibit better SER performance than the SM scheme at high Signal-to-Noise Ratios (SNR). Specifically, at an SER of $10^{-4}$ both the proposed schemes give an SNR gain of about 2.5 dB with respect to the SM scheme. However, the plot (b) in Fig. \ref{SERSM3BPCU} shows that the proposed scheme with High DoSM performs relatively better than the proposed scheme with Low DoSM at medium SNRs.

\begin{figure}
\centering
\includegraphics[scale=0.36]{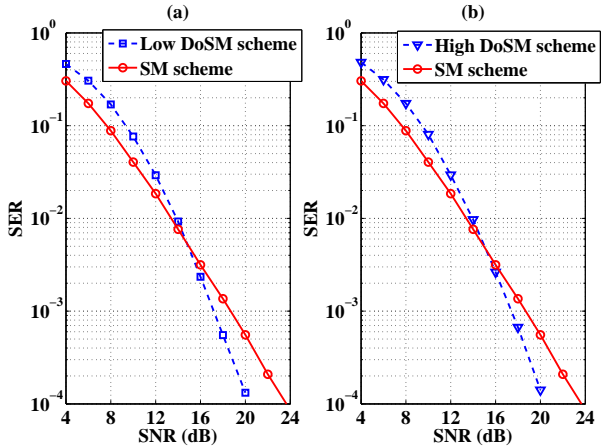}
\caption{Plots corresponding to that of Fig.{\ref{SERSM3BPCU}} for the spectral efficiency of 4 bpcu.}
\label{SERSM4BPCU}
\end{figure}

Fig. \ref{SERSM4BPCU} gives the SER curves of the proposed schemes and the SM scheme for the spectral efficiency of 4 bpcu. It is clear from the figure that the performance of both the proposed schemes are almost identical unlike the case of 3 bpcu. {\em Thus, indicating that the DoSM plays a key role in the performance of the SM scheme at low rates.} From the plots (a) and (b) of Fig. \ref{SERSM4BPCU}, it is evident that at an SER of about $10^{-4}$ both the proposed schemes give an SNR gain of about 3 dB with respect to the standard SM scheme.

Thus, for high rate applications, where $M$ is large compared to $N_t$, the proposed Low DoSM scheme is a suitable choice due to the observed constant coding gain property of the scheme. However, when $N_t$ is large or of the order of $M$, the proposed High DoSM scheme may give better SER performance than that of the Low DoSM scheme at low and medium SNRs.

\section{Conclusions}

Two modulation diversity techniques are proposed for the SM scheme using the CIOD meant for two transmit antennas. Both the proposed schemes use only one active antenna in any symbol duration and still achieve a transmit diversity order of two. Also, the proposed schemes are shown to admit low ML decoding complexity due the amicable structure of the CIOD.  It is observed through numerical simulations that the proposed scheme with Low DoSM has a constant coding gain for $M$-QAM signal sets with $M\geq16$. It is shown with our simulation results that both the proposed schemes outperform the SM scheme at medium and high SNRs due to higher diversity order.

\bibliographystyle{ieeetr}

\end{document}